\documentclass[aps,twocolumn,floatfix,eqsecnum,showpacs,prb,superscriptaddress]{revtex4}
\usepackage{graphicx}
\usepackage{amsmath}
\usepackage{amssymb}
\usepackage{color}
\usepackage{bm}
\usepackage{bigstrut}

\begin{document}
\title{Networks of quantum wire junctions: a system with
  quantized integer Hall resistance without vanishing longitudinal
  resistivity}

\author{Jaime Medina}
\affiliation{Facultad de Ciencias, Universidad Aut\'onoma de Madrid, 28049 Cantoblanco, Madrid, Spain}
\affiliation{Physics Department, Boston University, Boston, MA 02215, USA}
\author{Dmitry Green\footnote{dmitrygreen2009@gmail.com}}
\affiliation{170 East 83rd Street, New York, NY 10028, USA}
\author{Claudio Chamon}
\affiliation{Physics Department, Boston University, Boston, MA 02215, USA}

\date{\today}

\begin{abstract}
  We consider a honeycomb network built of quantum wires, with each
  node of the network having a Y-junction of three wires with a ring
  through which flux can be inserted. The junctions are the basic
  circuit elements for the network, and they are characterized by
  $3\times 3$ conductance tensors. The low energy stable fixed point
  tensor conductances result from quantum effects, and are determined by
  the strength of the interactions in each wire and the magnetic flux
  through the ring. We consider the limit where there is decoherence
  in the wires between any two nodes, and study the array as a network
  of classical 3-lead circuit elements whose characteristic
  conductance tensors are determined by the quantum fixed point. We
  show that this network has some remarkable transport properties in a
  range of interaction parameters: it has a Hall resistance quantized
  at $R_{xy}=h/e^2$, although the longitudinal resistivity is
  non-vanishing. We show that these results are robust against
  disorder, in this case non-homogeneous interaction parameters $g$ for
  the different wires in the network.
\end{abstract}

\maketitle


\section{Introduction}\label{introduction}

The transport properties of junctions of quantum wires are of interest
both seen from basic and applied perspectives. From the basic physics
aspect, quantum wires provide experimentally realizable ways for
studying interacting electrons in one-dimensional geometries, and in
particular junctions where 3 or more wires meet can display rather
rich behaviors. Theoretically, the problem of quantum wire junctions
is related to dissipative quantum mechanics in two or higher
dimensions, and to boundary conformal field
theory.~\cite{Chamon03,Oshikawa06} It also has a mathematical
connection to certain aspects of open string theory in a background
magnetic field.~\cite{Callan-etal,Callan-Freed} From a practical
viewpoint, junctions of quantum wires should serve as important
building blocks for the integration of quantum circuits, as they are
the natural element to split electric signals and serve as
interconnects.

Junctions of quantum wires have been the subject of many recent
studies,~\cite{Nayak99,Lal02,Chamon03,Chen02,Pham03,Rao04,Kazymyrenko05,Oshikawa06,Bellazzini07,Hou08,Das08,Agarwal_Das_Rao_Sen09,Bellazzini09a,Bellazzini09b,Safi09,Aristov10,Mintchev11,Aristov11,Aristov_Wolfe11,Wang_Feldman11,Caudrelier12}
which have uncovered many interesting transport properties as function
of interaction strength. Quantum wires with few transport channels, at
low energies, can be described as Tomonaga-Luttinger liquids,
characterized by a Luttinger parameter $g$ which encodes the
electron-electron
interactions.~\cite{Tomonaga50,Luttinger63,Mattis65,Haldane81} The
transport properties of a given junction depends on the Luttinger
parameters for each wire. At low energies, the conductance properties
of the junctions of $n$ wires are encoded in an $n\times n$
conductance tensor or matrix $G_{jk}$ that relate the incoming
currents to the applied voltages on the wires via $I_j=\sum_k
G_{jk}\,V_k$. At low voltages and low temperatures, the tensor takes
universal forms dictated by the nature of the infrared stable fixed
points in the renormalization group (RG) sense. These fixed points
have been categorized for the case of Y-junctions ($n=3$) of
spinless~\cite{Oshikawa06} and spinful~\cite{Hou08} electrons as
function of the interaction parameter $g$ when all the wires are
identical, and more recently in the case when the wires are not
identical and have different values $g_i$.~\cite{Hou12}

\begin{figure}[htbp]
\begin{center}
\includegraphics[width=0.345\textwidth]{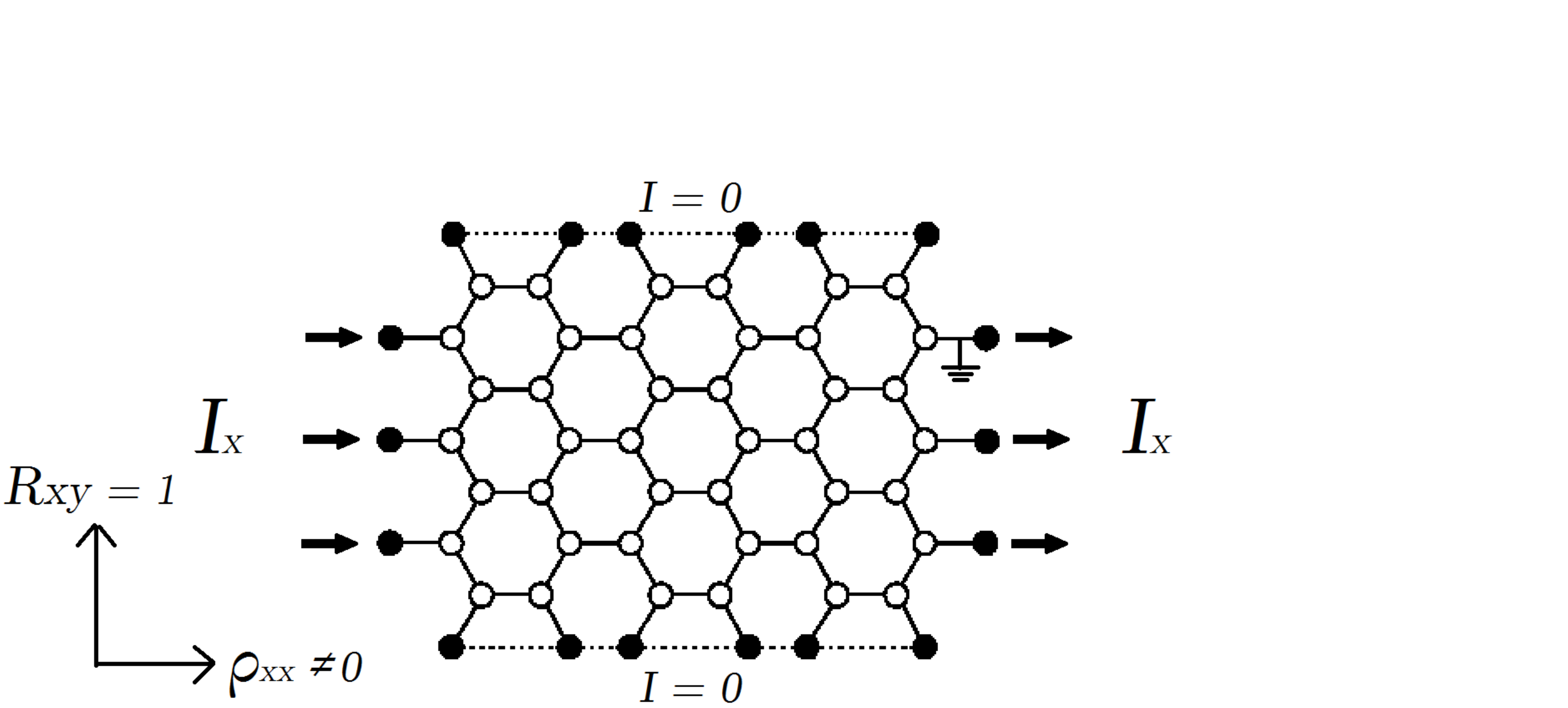}
\;\
\includegraphics[width=0.125\textwidth]{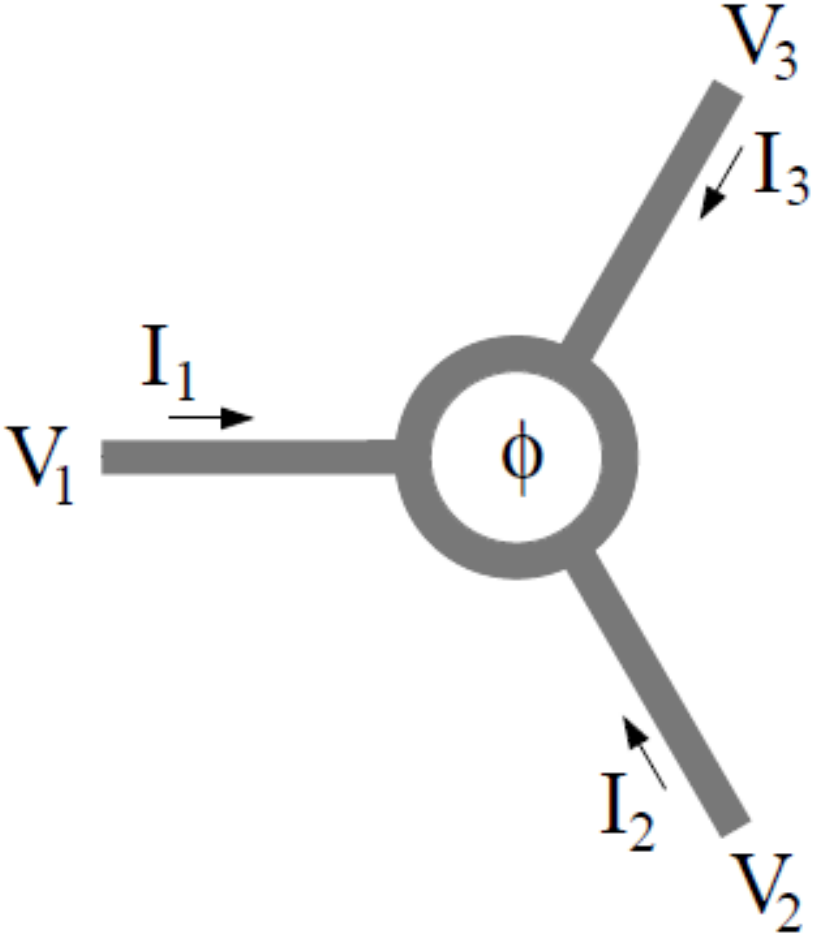}
\caption{(a) Scheme of a grid showing the flow of the current and the
  boundary conditions. External currents are fixed, as well as the
  potential on the node on the upper right corner. (b) Building block
  of the grid: junction of three quantum wires with a magnetic flux
  threading the ring. The $V_{1,2,3}$ are the voltages applied on each
  wire, and the $I_{1,2,3}$ the currents arriving at the junction from
  each of the three wires.}
\label{Fig1}
\end{center}
\end{figure}

In this paper we investigate the transport properties of networks
constructed using Y-junctions of quantum wires as building
blocks. Fig.~\ref{Fig1}a depicts an example of a network shaped in the
form of a rectangle, and Fig.~\ref{Fig1}b shows the individual
Y-junctions used in each node. We consider a simplified model where
the $3\times 3$ conductance tensor for each Y-junction is taken to be
that dictated by the low energy quantum RG fixed point, but the
transport is treated classically between any two junctions. The
treatment is sensible if the segment of the wires between two
junctions is large compared to the characteristic dephasing length in
the system. But the length scales of the junction itself, for example
the size of a ring as shown in Fig.~\ref{Fig1}b, should be smaller
than the dephasing length so that the junction is treated quantum
mechanically. The case when the full system is treated quantum
mechanically is extremely difficult to analyze, because it is an
interacting problem. For instance, a lattice version of the problem
would essentially be an example of a two-dimensional interacting
lattice model with a fermion sign problem.

We find rather remarkable results for the transport characteristics of
the network of Y-junctions, even when the role of quantum mechanics is
just to select the RG stable fixed point conductances of the
elementary building blocks. When the conductance is controlled by the
chiral fixed points $\chi_\pm$,~\cite{Chamon03,Oshikawa06} we find
that the whole network behaves as a Hall bar, with a Hall resistance
that is quantized to $R_{xy}=\pm h/e^2$, like in the integer quantum
Hall effect, with the sign given by the particular chirality of the
fixed points $\chi_+$ or $\chi_-$. However, the longitudinal
resistivity $\rho_{xx}\ne 0$, unlike in the case of the quantized
Hall effect where $\rho_{xx}$ vanishes. The quantization of $R_{xy}$
is a manifestation of the universal fixed point conductances. The
chiral fixed points are stable for a range of Luttinger parameters
$1<g<3$, and which of $\chi_+$ or $\chi_-$ is selected depends on the
flux threading the ring in the Y-junction.~\cite{Chamon03,Oshikawa06}
The flux breaks time-reversal symmetry, but it does not need to be
quantized at any given value; because of interactions, the conductance
of the Y-junction flows to fixed point values for a range of fluxes.

The quantization of $R_{xy}=\pm h/e^2$ for the network as a whole is
independent of the value of $g$ in the wires, as long as they are in
the range of stability of the chiral fixed points. Moreover, we show
that the quantization $R_{xy}=\pm h/e^2$ is stable against disorder in
the wire parameters. Specifically, we show that the quantization of
$R_{xy}$ remains even when the values of $g$ for different wires are not
uniform but disordered, {\it i.e.}, they are randomly distributed
around some average value $g$ with some spread $\delta g$.

The paper is organized as follows. In Sec.~\ref{model} we briefly
review the results for the conductance characteristics of single
quantum Y-junctions, which are the elementary building blocks for the
honeycomb wire-networks. In Sec.~\ref{analytical} we present
analytical results from which one can understand the origin of the
quantization of $R_{xy}$ when the conductance tensor of each of the
Y-junctions in the network is associated to a chiral fixed point. In
Sec.~\ref{numerical-results} we present numerical studies confirming the
analytical findings by analyzing grids with different values of the
interaction parameter $g$, different geometries and sizes, and
extrapolate these results to the thermodynamic limit. These numerical
calculations are of much value for the next step, taken in
Sec.~~\ref{disorder}, where we discuss the robustness of the
quantization of the Hall resistance in the case when the wires each
have different Luttinger parameters distributed randomly. The Appendix
contains a detailed description of the numerical method to
solve our network of Y-junctions.


\section{Single Y-junction as elementary circuit element}\label{model}

Each of these Y-junctions in the network consists of three
wires that are connected to a ring which can be threaded by a magnetic
flux, as shown in Fig. \ref{Fig1}b. This flux breaks time-reversal
symmetry, and the currents in the junction will depend on the
potential at its extremes and the magnetic flux inside the junction.

The current-voltage response of each Y-junction is determined by its
conductance tensor $G_{j k}$. Within linear response theory, the
total current $I_{j}$ flowing into the junction from wire $j$ is
related to the voltage $V_{k}$ applied to wire $k$ by
\begin{equation}\label{eq3}
I_{j}=\sum_{k}G_{j k}V_{k}
\end{equation}
where $j, k = 1, 2, 3$. Two sum rules apply to the conductance tensor
because of conservation of current and because the currents are
unchanged if the voltages are all shifted by a constant:
\begin{equation}
\sum_{j}G_{j k} = \sum_{k}G_{j k} = 0
\;.
\end{equation}

The $G_{jk}$ reach universal values at low temperatures and low bias
voltages. These universal values are dictated by the RG stable fixed
point that is reached for given values of the Luttinger parameters in
the wires. Here we shall focus on the case where all the three wires have
the same parameter $g$. In Sec.~\ref{disorder} we will consider the
more general case of network of wires where the three wires for each
Y-junction have different $g$'s.

When the three wires have the same $g$, the fixed point conductance tensor
has a $Z_{3}$ symmetry and takes the form~\cite{Oshikawa06}
\begin{equation}
G_{j k}=\frac{G_{S}}{2}(3\delta_{j k}-1)+\frac{G_{A}}{2}\epsilon_{j k}
\;,
\label{eq:G-GS-GA}
\end{equation}
where $\epsilon_{ij} =\delta_{i,j- 1}-\delta_{i,j+ 1}$ with $i+3\equiv
i$ and we separate the symmetric and anti-symmetric components of the
tensor, whose magnitudes are encoded in the scalar conductances
$G_{S}$ and $G_{A}$. $G_{A}$ vanishes when time-reversal symmetry is
not broken, for instance in the absence of magnetic flux through the
ring.

The fixed point values of $G_{S}$ and $G_{A}$ depend on the strength
of electron-electron interactions, encoded in the Luttinger parameter
$g$. We will focus on the chiral fixed points $\chi_{\pm}$, which are
stable in the range $1<g<3$.~\cite{Chamon03,Oshikawa06} In the chiral
cases, the conductances are given by
$G_{S}=G_{\chi}=\frac{e^{2}}{h}\frac{4g}{3+g^{2}}$ and $G_{A}=\pm g
\,G_{\chi}$. 
Thus the chiral conductance tensors
are:
\begin{equation}
G_{j k}^{\pm}=\frac{G_{\chi}}{2}[(3\delta_{j k}-1)\pm g\epsilon_{j k}]
\;.
\label{eq:chiral-G}
\end{equation}
We shall work in units where the quantum of conductance $e^2/h$ is set
to 1. 
 
The Y-junctions are then assembled into a network as shown in
Fig.~\ref{Fig1}a. We consider a regular hexagonal grid of Y-junctions
with $2c$ external connections on both the top and bottom sides and
$r$ on both the right and left side. Parametrized in such a way and
with wires of unit length, the dimensions of the grid as a function of
$r$ and $c$ are
\begin{eqnarray}
&& L_x = 6 c \nonumber \\
&& L_y = \sqrt{3} (2r +1)
\;.
\end{eqnarray}

In this grid we shall fix the
current flow along the $x$-axis from left to right and we shall fix the
currents flowing into the top and the bottom to zero, as shown in
Fig. \ref{Fig1}a. Given the conductance tensors at every node of the
network, we compute the potentials and the currents on the links of
the grid. The resistances and resistivities of the networks are
studied for different orientations and systems sizes, and for
different values of $g$. In appendix \ref{App3} we present details of
the method used to numerically compute the response of the networks.


\section{Analytical Results}\label{analytical}

We will measure the longitudinal and transverse responses in the
framework of the classical Hall problem by injecting a transverse
current along the $x$-axis and imposing a zero current boundary
condition along the two edges that are parallel to the $x$-axis. This
approach suggests that we solve for the potential in the bulk as a
function of the external current. In other words we need to invert
the fundamental equation \eqref{eq3} for $I$ and $V$ for each junction in
the bulk.

\begin{figure}[htbp]
\begin{center}
\includegraphics[width=0.5\textwidth]{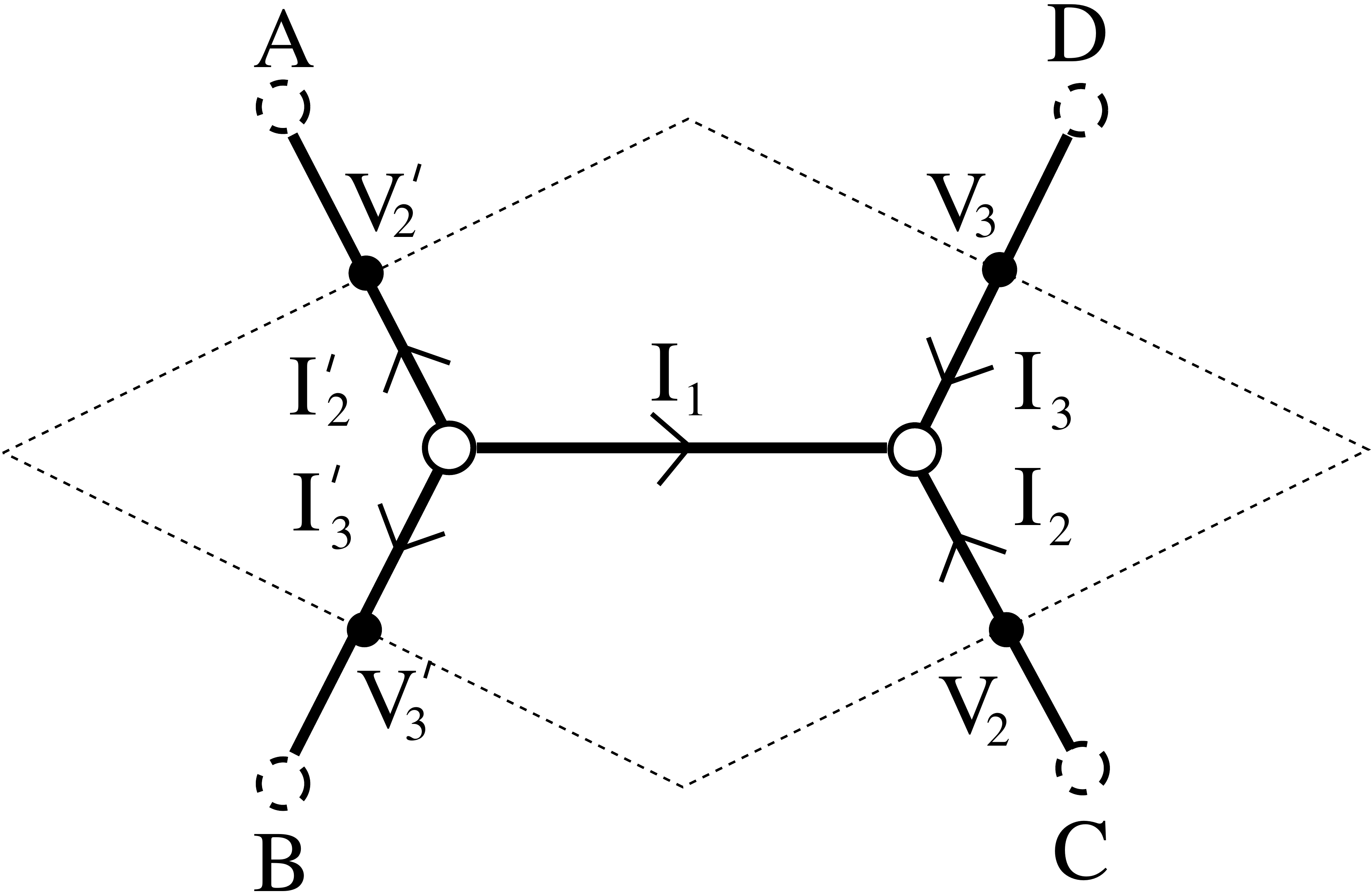}
\caption{Unit cell of the hexagonal network. Currents are assumed to
  be positive when directed along the arrows in the wires. Dotted
  lines denote the boundary of the unit cell. The rectangular region
  $ABCD$ shown is used for computing the resistances and
  resistivities of the network.}
\label{UnitCell}
\end{center}
\end{figure}

While the full network problem is not tractable analytically, we can
still gain some insight from a combination of analytics and
heuristics. In particular we will be able to prove quantization of the
transverse resistivity analytically, even with some forms of
disorder. Similarly we will derive the general form of the
longitudinal resistivity. We will confirm these results numerically in
later sections. Let us start with the unit cell of the hexagonal
lattice. There are two vertices (nodes) in each cell and current is
directed along the bonds (wires) as shown in Fig.~\ref{UnitCell}.
Looking at the right-hand node first, the potentials on the external
wires, $V_2$ and $V_3$, and the potential on the internal wire $V_1$
are defined only up to an additive constant. This means that
Eq. \eqref{eq3} is not invertible. However, by setting $V_1=0$, or
equivalently shifting all potentials in the two nodes by a constant
$V_i\rightarrow V_i-V_1$, the gauge is fixed and we obtain, using
Eq.~(\ref{eq:chiral-G}), the following:
\begin{equation}
\begin{pmatrix}
V_2-V_1\\
V_3-V_1
\end{pmatrix}
=\frac{1}{2g}
\begin{pmatrix}
2 & 1\mp g\\
1\pm g & 2
\end{pmatrix}
\begin{pmatrix}
I_2\\
I_3
\end{pmatrix}
\;.
\label{RightNode}
\end{equation}
The solution in the left node is similar but with the permutation
$(V_2,V_3)\rightarrow (V^\prime_2,V^\prime_3)$ and
$(I_2,I_3)\rightarrow -(I^\prime_2,I^\prime_3)$, which follows from
rotational symmetry and the orientation that we have chosen for the
currents. 

Now consider the potential gradient in the $x-$ and $y-$directions. It
is straightforward to derive the change in potential per unit cell,
$\Delta V_x$ and $\Delta V_y$, directly from Eq.~\eqref{RightNode} as
follows:

\begin{eqnarray}
&&\Delta_x V = V^\prime_3-V_2 = \frac{1}{2g}\left[2 I_1 - I_2 - I^\prime_3 \mp g\left(I^\prime_2 - I_3\right)\right] \nonumber \\
&& \Delta_y V = V_2-V_3 = \frac{1}{2g}\left[\pm g I_1 + I_2 -I_3\right].
\label{GradV}
\end{eqnarray}

It is instructive to consider a simple case. We will generalize this
result below, but for now consider a uniform current in the bulk in
the $x-$direction (or ``armchair'' configuration to borrow
nomenclature from graphene). Each horizontal wire in each unit cell
has a current $I_1=I$. By symmetry the other wires split the current
equally: $I_2=I_3=I^\prime_2=I^\prime_3=-I/2$. This configuration
leads to a particularly simple potential gradient: $\Delta_x V =
3I/2g$ and $\Delta_y V = \pm I/2$.

The result for the resistances and resistivities are apparent after we
account for the geometric factors. Consider the rectangular region
$ABCD$ in Fig.~\ref{UnitCell}, with sides $d_{AB}=\sqrt{3}$ and
$d_{AD}=2$. In the transverse direction the width of the rectangle is
twice the distance between the midpoint of the wires (with currents
$I_2$ and $I_3$), and the voltage drop $V_{AB}=2\,\Delta_y V$. The
Hall resistance (which coincides with the Hall resistivity
$\rho_{xy}$) is therefore $R_{xy}=V_{AB}/I=2\,\Delta_y V/I = \pm
1$. In other words the Hall resistance is independent of $g$ and
quantized to unity!

Similarly, in the longitudinal direction the length of the rectangle
is 4/3 the distance between the midpoint of the wires (with currents
$I'_2$ and $I_3$), and $V_{AD}=4/3\,\Delta_x V$. The longitudinal
resistance is $R_{xx}=4/3\,\Delta_x V/I=2/g$. There is an additional
geometric factor in the longitudinal resistivity given by
$\rho_{xx}=(d_{AB}/d_{AD})\,R_{xx}$, and it is thus given by
$\rho_{xx}=\sqrt{3}/g$. Hence the resistivity is non-zero and there is
dissipation unlike in the standard quantum Hall effect.

Had we used an alternate (``zigzag'') configuration where the
transverse current is zero $I_1=0$ and the uniform current is in the
$y-$direction, $I_3=I^\prime_3=-I_2=-I^\prime_2=I$, we would have
found a similar result, \textit{i.e.}, that the resistance in the
$x-$direction is quantized to $R_{xy}=\pm 1$ while the resistivity in
the $y-$ direction is $\rho_{yy}=\sqrt{3}/g$.

We find this result both unexpected and remarkable. By taking the
classical conductivity limit for each wire we have allowed decoherence
along the wires. However we have preserved the quantum coherence on
each vertex, as the chiral relation Eq.~(\ref{eq:chiral-G}) is by
nature a consequence of quantum scattering. Nonetheless even after
relaxing a portion of the coherence, some element of quantization in
the thermodynamic limit has survived in the form of an integer
quantized Hall resistivity. On the other hand, decoherence has
destroyed the zero longitudinal resistivity of the quantum Hall
effect, and so we are left with a hybrid quantum-classical Hall
effect. Note also that the simple uniform solution above suggests
robustness against disorder, another element of the integer quantum
Hall effect. As the transverse gradient of $V$ is independent of $g$
in the uniform bulk, suppose that $g$ is allowed to vary slowly from
vertex to vertex, more slowly than the current. In this regime we
would expect quantization to persist, and indeed we will confirm that
numerically later in this paper.

We will substantiate the assumptions and findings above numerically in
the next section.


\section{Numerical results}\label{numerical-results}

In this section we shall present numerical results for the voltages
and currents in the wires of the network. These numerical studies
serve first as a check of the analytical results presented in the
previous section~\ref{analytical} for the case where all the
interaction parameters are the same for all wires. Second, and more
importantly, they serve as a stepping stone to the case of
non-homogeneous (disordered) interaction parameters in the wires,
which will be considered in Sec.~\ref{disorder}. The method used to
solve for the voltages and currents in the grid is presented in
Appendix~\ref{App3}.

Let us focus on the armchair layout of Fig.~\ref{Fig1}a (similar
results follow in the case of the zigzag case). Also, without loss of
generality, we consider below only the $\chi_+$ fixed point. Current
is injected and collected uniformly into the wires on the right and on
the left of the network, respectively. More precisely, there are $r$
wires serving as connections to the outside on each side of the grid,
and current $I=I_x/r$ is injected in and collected out of these
external wires. The total current flowing along the horizontal or
$x$-direction is therefore $I_x$.

\begin{figure}[htbp]
\begin{center}
\includegraphics[width=0.5\textwidth]{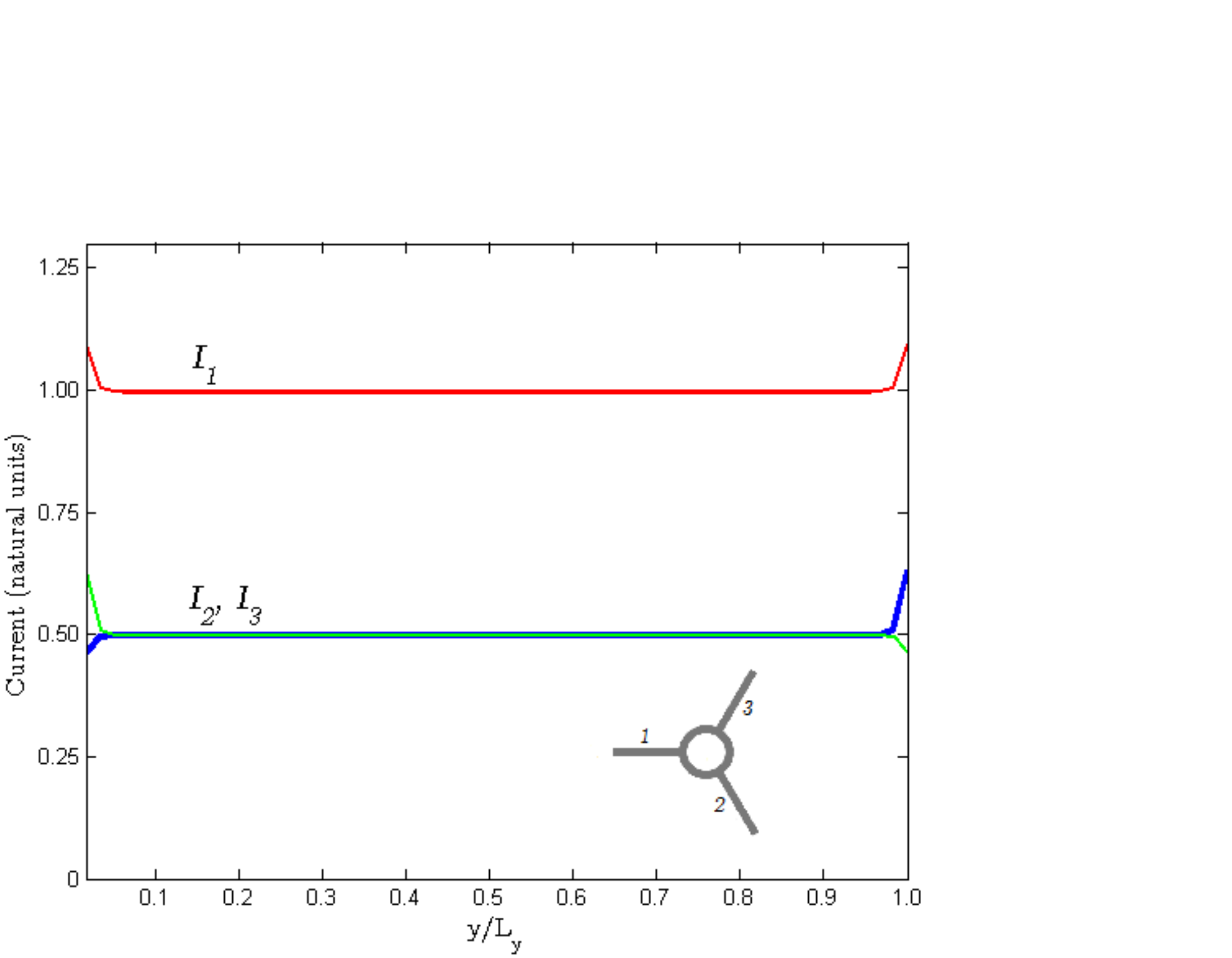}
\caption{Currents flowing through the Y-junctions that lie along a
  vertical line in the middle of the bar ($x=L_x/2$) as a function of
  vertical position $y/L_y$. Note that for $y$ values away from the
  edges the currents tend to $I_1=1$ and $I_{2,3}=1/2$, as predicted
  analytically for the asymptotic limit.}
\label{Fig5}
\end{center}
\end{figure}

The distribution of the currents in the inner parts of the grid that
follow from this uniform injection of external currents is shown in
Fig.~\ref{Fig5}. We find a close to uniform distribution, with
slightly larger currents closer to the edges. This distribution is
independent of the value of $g$. These patterns of current flow in the
inner wires of the grid are in agreement with the current
distributions discussed in the analytical studies of the previous
section.

The Hall voltage is the potential drop $V_y$ along the vertical or
$y$-direction. We note that the potential drop $V_y$ is computed by
looking at the potentials for two points at the same horizontal
position ({\it i.e.}, the same $x$ position), one at the top and one
at the bottom of the network.

\begin{figure}[htbp]
\begin{center}
\includegraphics[width=0.5\textwidth]{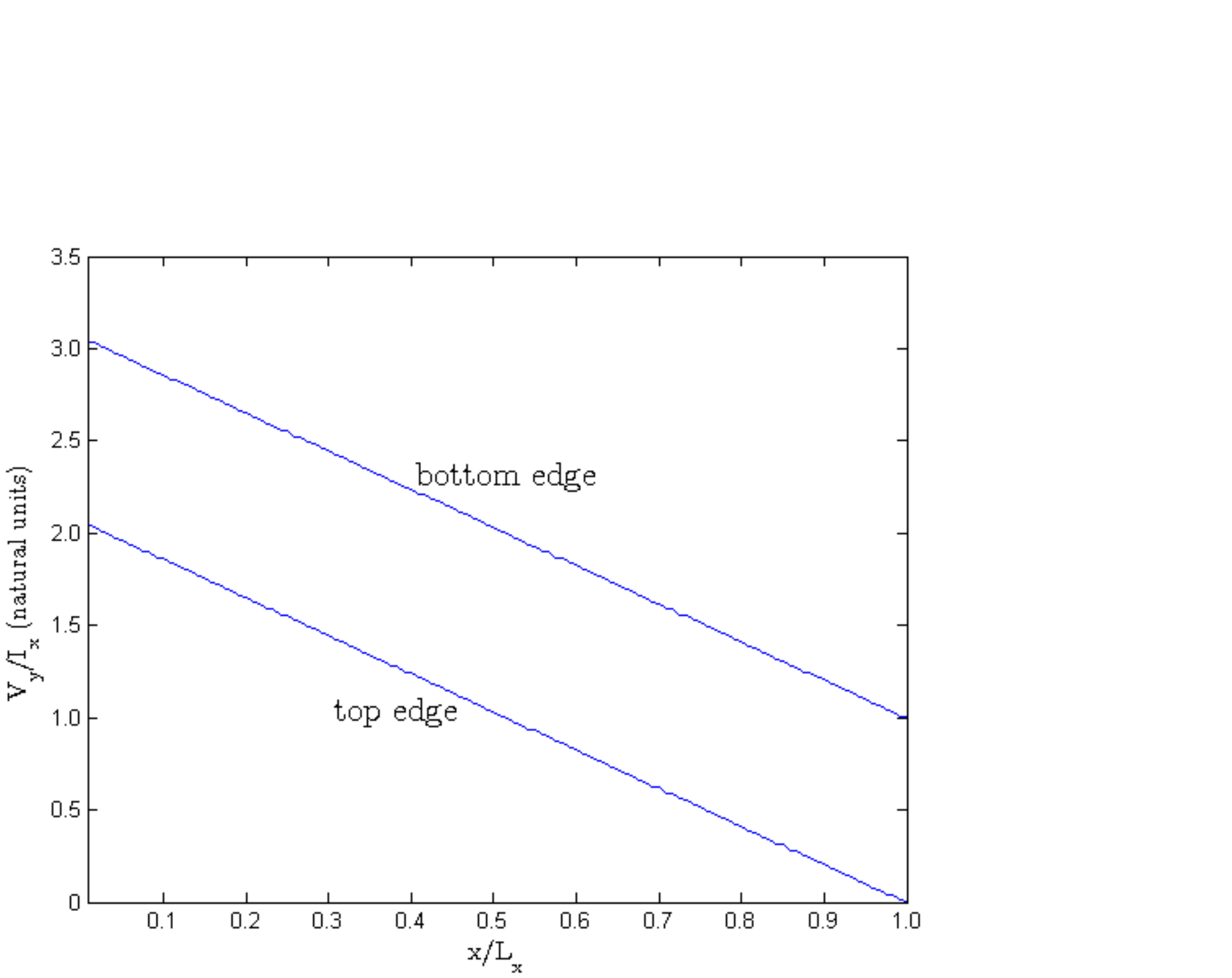}
\caption{Voltages at the top and and bottom edges as function of
  horizontal position $x/L_x$ when the node at the top right corner is
  grounded. The grid size is $r=50$, $c=60$ and $g=\sqrt{3}$. Notice that the
  difference between the voltages at the top and bottom edges for a
  given $x/L_x$ is exactly 1 in natural units.}
\label{Fig13}
\end{center}
\end{figure}

We show in Fig. \ref{Fig13} the potentials measured at the top and at
bottom of the (rectangular shaped) grid. Notice that the potentials
drop linearly with the horizontal direction, but that the difference
between the two potentials, $V_y$, is constant.

The Hall resistance is computed as follows. Let $\bar V_y$ be the
average over the horizontal positions $x$ of the Hall voltage
drop. (Since in this case without disorder $V_y$ is constant, the
average is actually unnecessary here.) Then the Hall resistance is
given by $R_{xy}=\bar V_y/I_x$. We find numerically that $R_{xy}=1$ as
expected from the analytical arguments.~\footnote{The value $R_{xy}=1$
  that is found numerically is exact to double precision in Matlab.}
Recall that we are working in units where $e^2/h=1$, so indeed we have
\begin{equation}
  \label{eq:hall-rxy}
R_{xy}=\frac{h}{e^2 }
\;,
\end{equation}
which we find is independent of the value of $g$. We remark that we
find that this quantization holds independent of the aspect ratio,
orientation (armchair {\it vs.} zigzag) or size of the grid. 

We also computed the potential difference between points on the left
and on the right sides of the grid, $V_x$, as a function of the
vertical direction $y$. In this case we find that the horizontal
potential difference is almost constant as function of $y$ (as opposed
to the case of the vertical drop $V_y$, which is exactly independent
of $x$). The difference is bigger, by an amount of order $1/L_y$, when
$y$ is in the middle of the grid as compared to when $y$ is at the
edges. We define $\bar V_x$ as the $y$-position averaged voltage
difference between the left and right sides of the grid. The
longitudinal resistance is given by $R_{xx}=\bar V_x/I_x$, and the
longitudinal resistivity by $\rho_{xx}=L_y/L_x\;\bar V_x/I_x$.

We find that the longitudinal resistance is non-zero, in agreement
with Sec.~\ref{analytical}. We find numerically, however, that there
are finite system size corrections to the analytical predictions. We
find that
\begin{equation}\label{eq2}
R_{xx}(g,L_x,L_y)=\frac{\sqrt{3}}{g} \frac{L_x}{L_y-A(L_x,L_y)}
\;,
\end{equation}
where $A$ is a factor of order 1 that corrects for finite sizes. We
find numerically that in the thermodynamic limit $A\to 1$ for the
armchair configuration, whereas $A=0$ independent of system size in
the zigzag case. Therefore, in the thermodynamic limit we obtain
\begin{equation}
\rho_{xx}=\lim_{L_x,L_y\to\infty}
\frac{L_y}{L_x}\;R_{xx}(g,L_x,L_y)=
\frac{\sqrt{3}}{g} 
\;,
\end{equation}
in agreement with the result in Sec.~\ref{analytical}.

\begin{figure}[htbp]
\begin{center}
\includegraphics[width=0.5\textwidth]{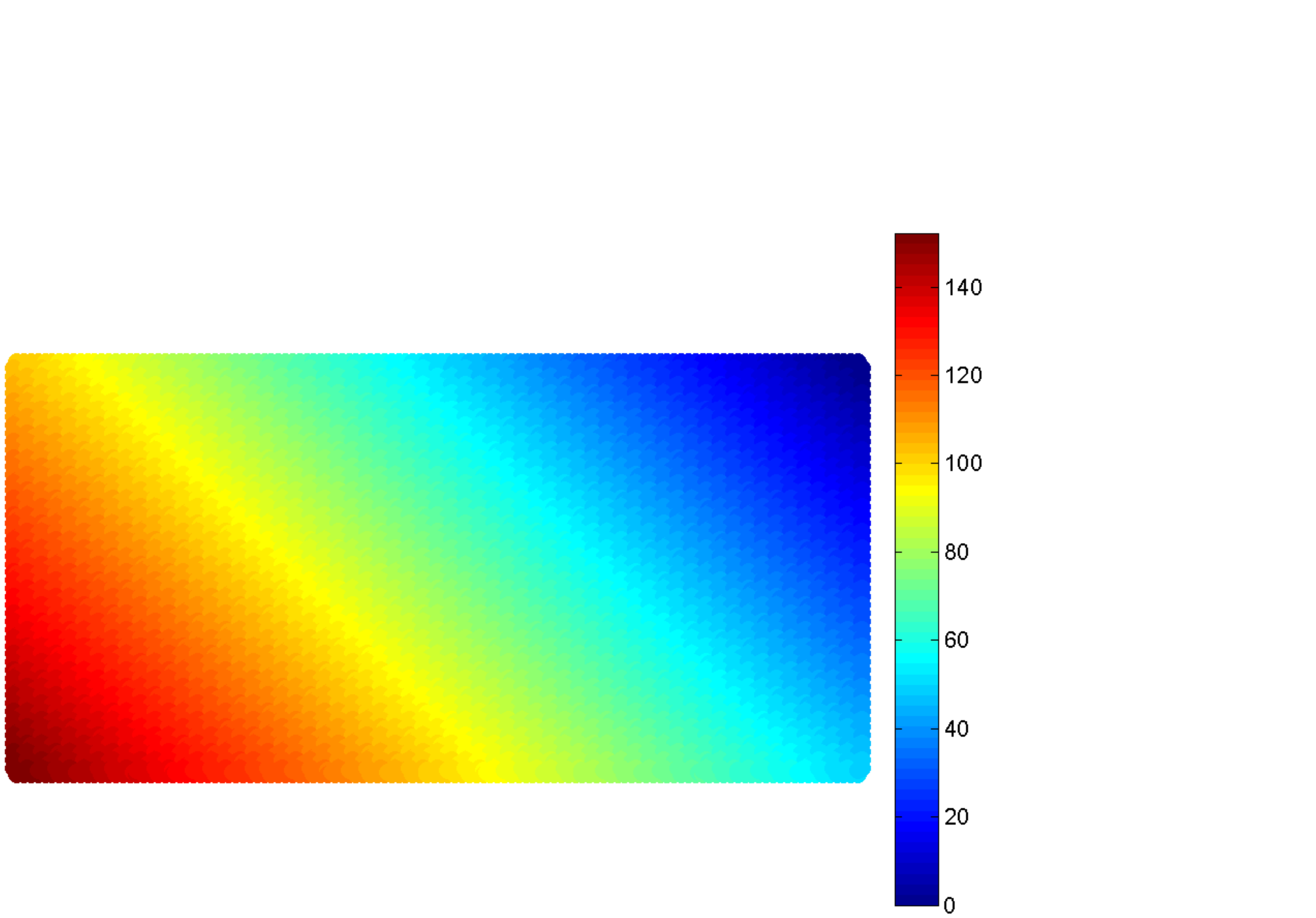}
\caption{Density plot for the voltages on the grid nodes, for a system
  with $r=50$, $c=60$ and $g=\sqrt{3}$. Notice the constant slope of
  the equipotential lines, which is related to the Hall angle
  $\theta_H$. The Hall angle depends on the interaction strength and
  is given by Eq.~\ref{eq:Hall_angle_g}.}
\label{Fig4}
\end{center}
\end{figure}

The Hall angle $\theta_H$ is given by $\tan
\theta_H=\rho_{xy}/\rho_{xx}$, and we naturally find, given the
agreement with the results for $\rho_{xx}$ and $\rho_{xy}$ above, that
\begin{equation}
\tan \theta_H = \frac{g}{\sqrt{3}}
\label{eq:Hall_angle_g}
\end{equation}
in the thermodynamic limit. This Hall angle can be visualized very
naturally by plotting the voltages at the wires on the grid, as shown
in Fig.~\ref{Fig4}. The Hall angle appears as the slope of the lines
of constant voltage. These equipotential lines are straight in this
example where all the wires have the same interaction parameter $g$;
this is no longer the case when disorder is introduced in the next
Sec.~\ref{disorder}.

\section{Robustness against disorder}\label{disorder}

In this section we will generalize the wire networks to the case when
the interaction parameters $g$ for each of the wires in the network
are not uniform, but instead are drawn independently from a
distribution. We shall consider a distribution in which $g$ in each of
the wires in the network takes a value between $(\bar{g}-\delta g,
\bar{g}+\delta g)$, with uniform probability. Because the interaction
parameter should be positive, $\delta g < \bar{g}$.

When the wires connecting to a given Y-junction have different values
of $g$, the conductance tensor $G_{ij}$ for a chiral fixed point is no
longer given by Eq.~(\ref{eq:chiral-G}), but instead it takes the form (see
Ref.~\onlinecite{Hou12})
\begin{equation}
G_{jk} = 2 \frac{g_j (g_1 + g_2 + g_3) \delta_{jk} + g_j g_k (\pm g_m \epsilon_{jkm}-1)}{g_1 g_2 g_3 +g_1 +g_2 +g_3}
\;.
\end{equation}
Using this conductance tensor, one can compute numerically
(using the method of Appendix~\ref{App3}) the voltages and currents in
all wires of the network for a given realization of the disorder.

We shall show below that the quantization $R_{xy}=1$ of the Hall
conductance that we found in the clean limit remains , in the
thermodynamic limit, even in the presence of disorder. For a finite
lattice, as one should expect, there are fluctuations that we quantify
below for the armchair configuration.

\begin{figure}[htbp]
\begin{center}
\includegraphics[width=0.5\textwidth]{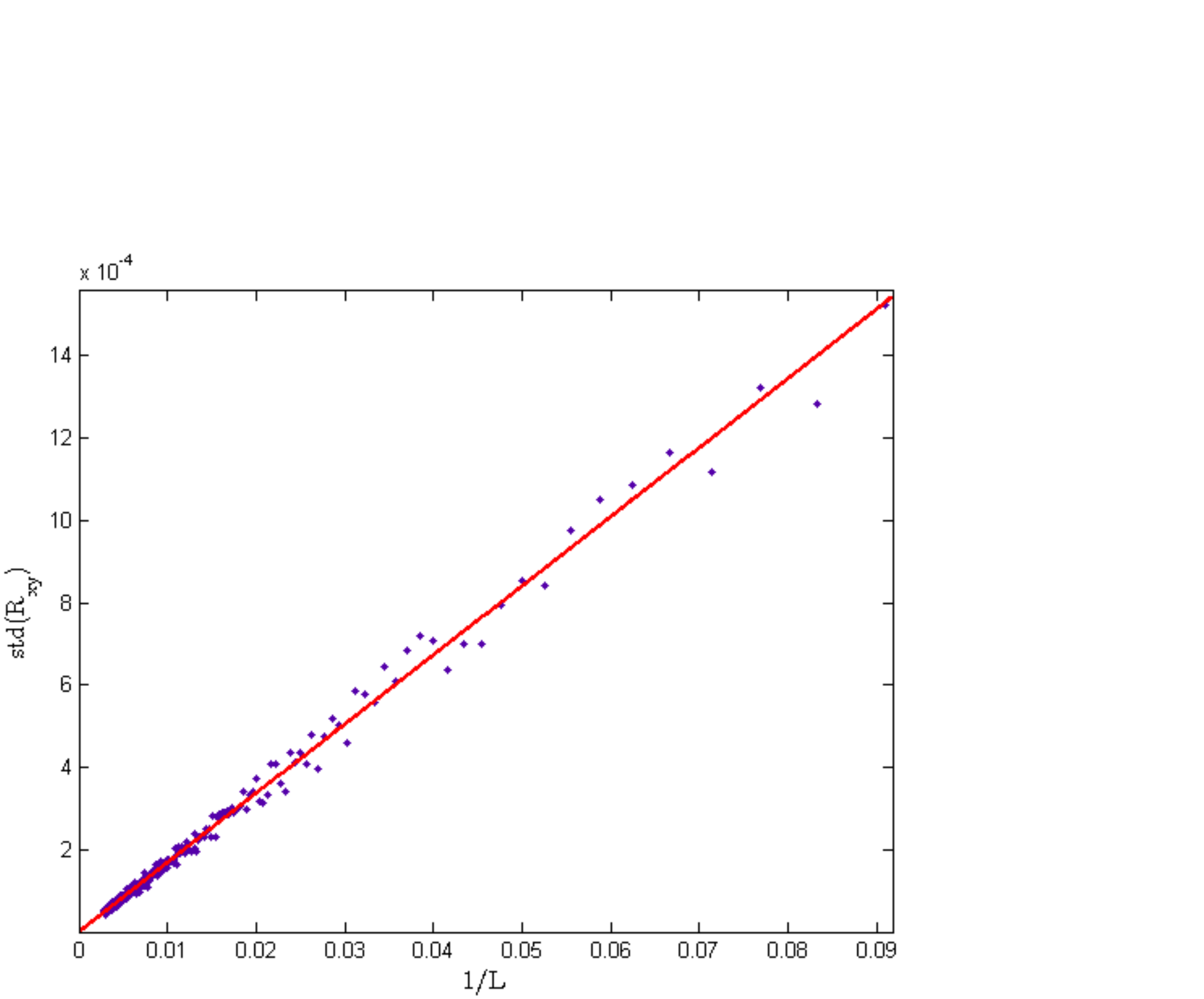}
\caption{Standard deviation of the Hall resistance for 100 simulations
  with $\bar{g}=\sqrt{3}$ and $\delta g=\bar{g}/10$ as a function of $1/L$ for a
  grid with $r=c=L$ (which fixes the aspect ratio). It scales to
  zero in the large $L$ limit, implying that the system is
  self-averaging and the Hall resistance $R_{xy}\to 1$ independent of
  disorder in the thermodynamic limit.}
\label{Fig9}
\end{center}
\end{figure}

We compute Hall resistance $R_{xy}$ (defined as the average of the
voltage differences between top and bottom of the network, divided by
the injected current) for several realizations of disorder and system
sizes. For a fixed system size, we then find the disorder average
$\overline R_{xy}$ and standard deviation $\Delta R_{xy}=
\sqrt{\overline {R^2_{xy}}-{\overline R_{xy}}^2}$ of $R_{xy}$.  We
find that $\overline R_{xy}\to 1$ as the number of realizations
increase, and that the standard deviation $\Delta R_{xy}\to 0$ as
$L$ increases (we use lattices with $r=c=L$). We show in
Fig. \ref{Fig9} the finite size scaling of the $\Delta R_{xy}$. That
$\Delta R_{xy}\to 0$ in the thermodynamic limit means that the system
is self-averaging, and therefore $R_{xy}\to 1$ independent
of disorder in the thermodynamic limit. We conclude then that
quantization is robust against disorder.

We have also checked the effects of disorder for the zigzag
configuration, reaching similar conclusions that disorder does not
alter the quantization of the conductance in the thermodynamic limit.

In summary, we find that, in the thermodynamic limit, the general
results of the previous sections hold even in the presence of
disorder.


\section{Conclusions}

We investigated the transport properties of hexagonal networks whose
nodes are Y-junctions of quantum wires. In our model the $3\times 3$
conductance tensor for each Y-junction is dictated by the low energy
RG fixed point, but the transport is treated classically between any
two junctions. We find a surprising result: in spite of relaxing
quantum coherence between the junctions, we find a quantized Hall
resistance.

Specifically, in the regime where the junction conductance is
controlled by the chiral fixed points
$\chi_\pm$,~\cite{Chamon03,Oshikawa06} (when the interaction parameter
obeys $1<g<3$), the network exhibits a quantized Hall resistance:
$R_{xy}=\pm h/e^2$. This quantization is similar to that in the
integer quantum Hall effect. Further, the quantization is independent
of the interaction parameter $g$ even in the presence of disorder in
$g$. The quantization of the Hall resistance follows from the specific
form of the conductance tensor at the RG stable chiral fixed point at
each Y-junction. However, unlike in the quantized Hall effect, where
the longitudinal resistivity vanishes, $\rho_{xx}$ is not zero:
$\rho_{xx}=(\sqrt{3}/g)\,h/e^2$. Dissipation in the longitudinal
direction is a result of decoherence within the wires. We emphasize
that in our model the wires are classical, but the nodes remain
quantum mechanical and the form of the conductance tensor $G$ at each
junction is constrained by quantum scattering effects. The essential
ingredient for the quantization of the Hall conductance is the value
of the chiral fixed point conductance of the individual junctions.

Finally, let us comment on the finite temperature corrections to the
value $R_{xy}=\pm h/e^2$ in the network. As opposed to the case of the
quantum Hall effect where the quantization is exponentially accurate
because of an energy gap, the quantization in the networks has a power
law correction in $T$ because the wire networks are gapless. The
quantization should be as accurate as the conductance tensor is close
to that of the RG fixed point. The corrections to the conductance
tensor scale as $T^\Delta$, where $\Delta=4g/(3+g^2)$ is the scaling
dimension of the leading irrelevant operator at the chiral fixed
points.~\cite{Chamon03,Oshikawa06}

Notice that the temperature scaling of the conductivity above should
hold only under the assumption of decoherence within the
wires. However, as temperature goes to zero, the coherence length
increases, and therefore there is an implicit assumption of order of
limits for the results in this paper to work as presented: the length
of the wires should be taken to infinity before the limit of $T=0$ is
taken. But it is natural to wonder whether the quantization that we
found in this work should persist or not even if transport along the
wires is always coherent. Indeed, one possibility is that in the
coherent regime one might have quantization of the Hall conductance
with vanishing longitudinal resistivity. However, to address this
regime one would need to tackle the fully interacting two-dimensional
fermionic model, which is beyond the scope of this paper. One route to
follow could be to consider a lattice model where the wires are
described by a tight binding model, with three wires coupled together
at junctions by hopping matrix elements between them. One could
possibly start with a non-interacting version of the model, where the
chiral conductances used in this paper are obtained by fine tuning to
the fixed point (since the non-interacting model is marginal and there
is no RG flow). The problem then becomes one of electrons in a
superlattice, with the number of bands scaling with the number of
sites describing the wires within a supercell. The Hall conductance
for this tight-binding model could be obtained by computing the Chern
number of the filled bands. If the Hall conductance does not vanish in
this model, it is only protected algebraically in temperature, as
there would be ``mini gaps'' separating bands that scale inversely
with the size of the wires, instead of true band gaps. Analyzing such
model may shine some light on the problem of wire networks in the
coherent regimes.

\section*{Acknowledgments}
This work was supported in part by the DOE Grant No. DE-FG02-
06ER46316 (C.C.).

\appendix

\section{Method}\label{App3}

Our numerical approach consists of solving the full lattice model
exactly. In this section we describe our methodology in detail.

\begin{figure}[htbp]
\begin{center}
\includegraphics[width=0.5\textwidth]{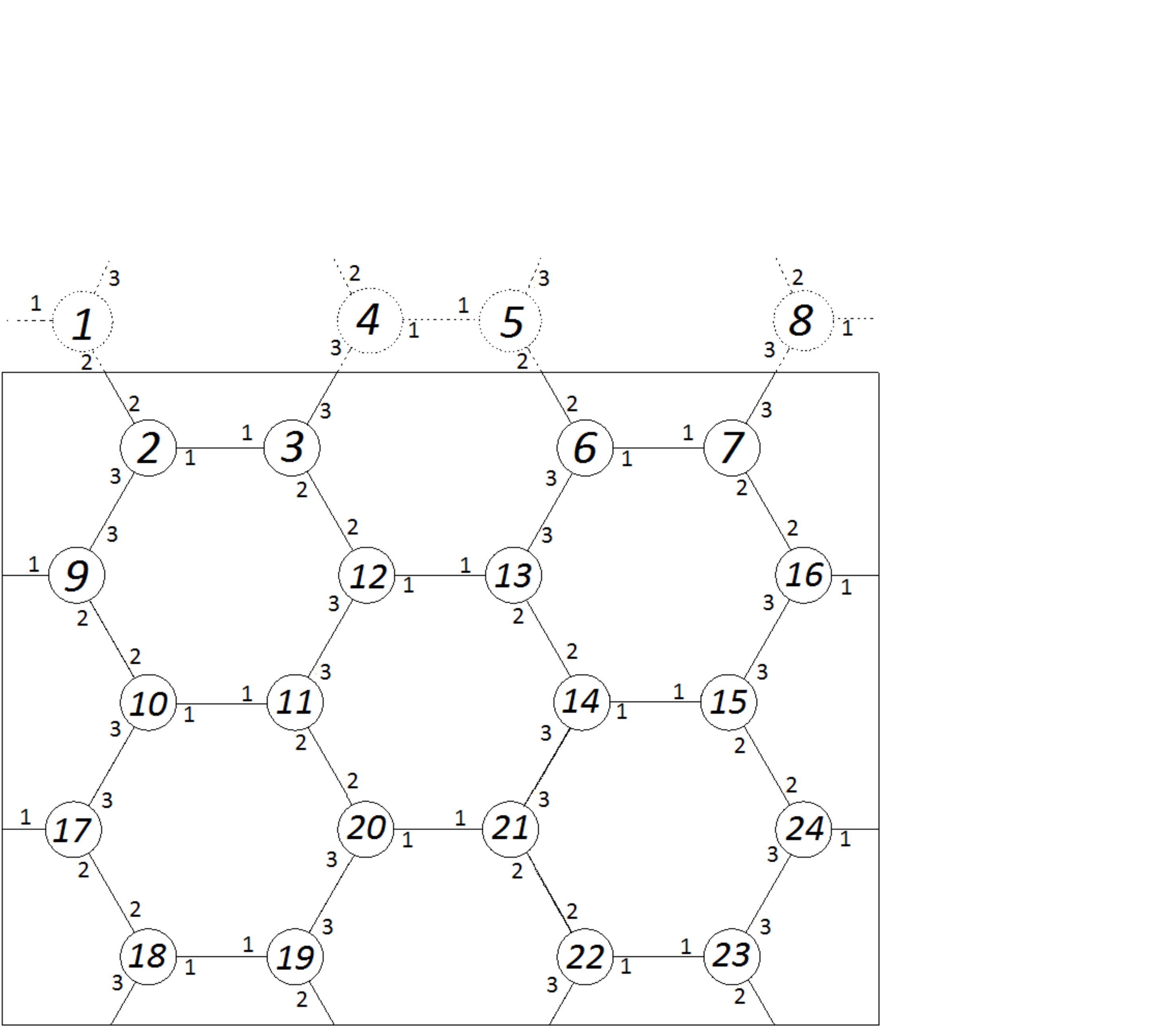}
\caption{Example with $r=2$, $c=2$. Note the row of ``ghost nodes'' at
  the top edge.}
\label{Fig3}
\end{center}
\end{figure}

Consider an arbitrary lattice with $r$ external wires on each side and
$2c$ external wires at the top and bottom edges. An equal current
will be injected into each of the $r$ wires on the left, and the $2c$
edge wires will have a current of zero. This defines the boundary
conditions. A $2\times 2$ lattice is shown for example in
Figure~\ref{Fig3}. For later convenience we include a row of $2c$
``ghost nodes'', shown as dotted lines at the top edge, but they are
only there to facilitate the numbering scheme and no current will flow
through them. Including the ghost nodes there are a total of
$N=4c(r+1)$ nodes.

The points on each wire that emanate from each node are governed by
the equation $V=GI$ where $G$ is a $3\times 3$ matrix. Thus we start
with $3N$ degrees of freedom. However, starting in this way
introduces many redundant variables in the bulk because in a classical
wire the current is the same everywhere along the wire and so is the
potential. We will unify the two points on each wire in the bulk by
imposing a set of constraints. In general there are $6cr+c-r$ such constraints,
which equals the number of wires in the bulk.

To write down the full network equation let us label each of the $3N$
points by $(n,i)$, where $n=1,\dots,N$ is the node index and $i=1,2,3$
refers to the point on each wire that emanates from each node. The
potentials and currents at each of these points are denoted by
$V^{(n)}_i$ and $I^{(n)}_i$, respectively. To illustrate this
notation, in Fig.~\ref{Fig3} the constraint along the wire that
connects nodes $6$ and $7$ would be written as $V^{(6)}_1=V^{(7)}_1$
and $I^{(6)}_1=-I^{(7)}_1$.

Each node obeys the relation $V^{(n)}=G^{(n)}I^{(n)}$ where $G^{(n)}$ is the $3\times 3$
matrix from Eq. \eqref{eq3}. The network is thus described by the
following linear equation with constraints:

\begin{equation}
\begin{pmatrix}
I^{(1)}\\
I^{(2)}\\
\vdots \\
I^{(N)}
\end{pmatrix}
=
\begin{pmatrix}
G^{(1)} & 0 & \cdots & 0\\
0 & G^{(2)} & \cdots & 0\\
\vdots & \vdots & \ddots & \vdots\\
0 & 0 & \cdots & G^{(N)}
\end{pmatrix}
\begin{pmatrix}
V^{(1)}\\
V^{(2)}\\
\vdots \\
V^{(N)}
\;
\label{NetworkEq}
\end{pmatrix}
\end{equation}

Next we impose the constraints to reduce the effective dimensionality of the problem. Start with the set of point pairs on each wire in the bulk $\{(n,i),(m,i)\}$, where $n$ and $m$ are nearest neighbor nodes. The constraints are $V^{(n)}_i=V^{(m)}_i$ and $I^{(n)}_i=-I^{(m)}_i$ for each pair. We impose the constraint on voltages by adding the $(n,i)$-th and $(m,i)$-th {\it columns} together, removing the $(m,i)$-th column and removing $V^{(m)}_i$ from the vector of potentials in Eq.~\eqref{NetworkEq}. Similarly we impose the constraint on currents by adding the $(n,i)$-th and $(m,i)$-th {\it rows}, deleting the $(m,i)$-th row and removing $I^{(m)}_i$ from the vector of currents. Also we replace the current $I^{(n)}_i$ that has not been eliminated by zero because $I^{(n)}_i+I^{(m)}_i=0$. Therefore each constraint is equivalent to removing one row and one column and reduces the dimensionality of the original problem by one. Furthermore, we have replaced each current in the bulk by zero which is important because the only currents that are left in Eq.~\eqref{NetworkEq} are fully determined, being equal to either zero in the bulk or to the boundary conditions. 

Eliminating the ghost nodes is straightforward -- we simply remove the ghost currents, potentials and their associated rows and columns in Eq.~\eqref{NetworkEq}. This reduces the dimensionality further by $3\times 2c$, which is the number of wires emanating from the ghost nodes. The final step is to fix the gauge. Since all potentials are determined up to an overall constant, we pick an arbitrary potential, set it to zero, and remove the associated row and column from Eq.~\eqref{NetworkEq}.

To summarize, we started with $3N=12c(r+1)$ redundant degrees of freedom and then through successive transformations we imposed $6cr+c-r$ constraints in the bulk, eliminated $6c$ ghost points, and fixed one potential to zero. The dimensionality has thus been reduced to $6rc+5c+r-1$ and, crucially, the only currents appearing are either zero or fixed by boundary conditions. Having eliminated all redundancies allows us to solve for the potential at any point, as a function of the boundary currents, by inverting the reduced version of Eq.~\eqref{NetworkEq}, which we do numerically.

The generalization to random couplings $g$ is straightforward. The derivation proceeds in exactly the same way as we just described, but we start with non-uniform $G^{(n)}$.


\bibliography{refer-junction}

\begin{thebibliography}{28}
\expandafter\ifx\csname natexlab\endcsname\relax\def\natexlab#1{#1}\fi
\expandafter\ifx\csname bibnamefont\endcsname\relax
  \def\bibnamefont#1{#1}\fi
\expandafter\ifx\csname bibfnamefont\endcsname\relax
  \def\bibfnamefont#1{#1}\fi
\expandafter\ifx\csname citenamefont\endcsname\relax
  \def\citenamefont#1{#1}\fi
\expandafter\ifx\csname url\endcsname\relax
  \def\url#1{\texttt{#1}}\fi
\expandafter\ifx\csname urlprefix\endcsname\relax\def\urlprefix{URL }\fi
\providecommand{\bibinfo}[2]{#2}
\providecommand{\eprint}[2][]{\url{#2}}

\bibitem[{\citenamefont{Chamon et~al.}(2003)\citenamefont{Chamon, Oshikawa, and
  Affleck}}]{Chamon03}
\bibinfo{author}{\bibfnamefont{C.}~\bibnamefont{Chamon}},
  \bibinfo{author}{\bibfnamefont{M.}~\bibnamefont{Oshikawa}}, \bibnamefont{and}
  \bibinfo{author}{\bibfnamefont{I.}~\bibnamefont{Affleck}},
  \bibinfo{journal}{Phys. Rev. Lett.} \textbf{\bibinfo{volume}{91}},
  \bibinfo{pages}{206403} (\bibinfo{year}{2003}).

\bibitem[{\citenamefont{Oshikawa et~al.}(2006)\citenamefont{Oshikawa, Chamon,
  and Affleck}}]{Oshikawa06}
\bibinfo{author}{\bibfnamefont{M.}~\bibnamefont{Oshikawa}},
  \bibinfo{author}{\bibfnamefont{C.}~\bibnamefont{Chamon}}, \bibnamefont{and}
  \bibinfo{author}{\bibfnamefont{I.}~\bibnamefont{Affleck}},
  \bibinfo{journal}{J. Stat. Mech.: Theory Exp.}
  \textbf{\bibinfo{volume}{2006}}, \bibinfo{pages}{P02008}
  (\bibinfo{year}{2006}).

\bibitem[{\citenamefont{Callan et~al.}(1995)\citenamefont{Callan, Klebanov,
  Maldacena, and Yegulalp}}]{Callan-etal}
\bibinfo{author}{\bibfnamefont{C.~G.} \bibnamefont{Callan}},
  \bibinfo{author}{\bibfnamefont{I.~R.} \bibnamefont{Klebanov}},
  \bibinfo{author}{\bibfnamefont{J.~M.} \bibnamefont{Maldacena}},
  \bibnamefont{and} \bibinfo{author}{\bibfnamefont{A.}~\bibnamefont{Yegulalp}},
  \bibinfo{journal}{Nuclear Physics B} \textbf{\bibinfo{volume}{443}},
  \bibinfo{pages}{444} (\bibinfo{year}{1995}).

\bibitem[{\citenamefont{Callan and Freed}(1992)}]{Callan-Freed}
\bibinfo{author}{\bibfnamefont{C.~G.} \bibnamefont{Callan}} \bibnamefont{and}
  \bibinfo{author}{\bibfnamefont{D.}~\bibnamefont{Freed}},
  \bibinfo{journal}{Nuclear Physics B} \textbf{\bibinfo{volume}{374}},
  \bibinfo{pages}{543 } (\bibinfo{year}{1992}).

\bibitem[{\citenamefont{Nayak et~al.}(1999)\citenamefont{Nayak, Fisher, Ludwig,
  and Lin}}]{Nayak99}
\bibinfo{author}{\bibfnamefont{C.}~\bibnamefont{Nayak}},
  \bibinfo{author}{\bibfnamefont{M.~P.~A.} \bibnamefont{Fisher}},
  \bibinfo{author}{\bibfnamefont{A.~W.~W.} \bibnamefont{Ludwig}},
  \bibnamefont{and} \bibinfo{author}{\bibfnamefont{H.~H.} \bibnamefont{Lin}},
  \bibinfo{journal}{Phys. Rev. B} \textbf{\bibinfo{volume}{59}},
  \bibinfo{pages}{15694} (\bibinfo{year}{1999}).

\bibitem[{\citenamefont{Lal et~al.}(2002)\citenamefont{Lal, Rao, and
  Sen}}]{Lal02}
\bibinfo{author}{\bibfnamefont{S.}~\bibnamefont{Lal}},
  \bibinfo{author}{\bibfnamefont{S.}~\bibnamefont{Rao}}, \bibnamefont{and}
  \bibinfo{author}{\bibfnamefont{D.}~\bibnamefont{Sen}},
  \bibinfo{journal}{Phys. Rev. B} \textbf{\bibinfo{volume}{66}},
  \bibinfo{pages}{165327} (\bibinfo{year}{2002}).

\bibitem[{\citenamefont{Chen et~al.}(2002)\citenamefont{Chen, Trauzettel, and
  Egger}}]{Chen02}
\bibinfo{author}{\bibfnamefont{S.}~\bibnamefont{Chen}},
  \bibinfo{author}{\bibfnamefont{B.}~\bibnamefont{Trauzettel}},
  \bibnamefont{and} \bibinfo{author}{\bibfnamefont{R.}~\bibnamefont{Egger}},
  \bibinfo{journal}{Phys. Rev. Lett.} \textbf{\bibinfo{volume}{89}},
  \bibinfo{pages}{226404} (\bibinfo{year}{2002}).

\bibitem[{\citenamefont{Pham et~al.}(2003)\citenamefont{Pham, Pi\'echon, Imura,
  and Lederer}}]{Pham03}
\bibinfo{author}{\bibfnamefont{K.-V.} \bibnamefont{Pham}},
  \bibinfo{author}{\bibfnamefont{F.}~\bibnamefont{Pi\'echon}},
  \bibinfo{author}{\bibfnamefont{K.-I.} \bibnamefont{Imura}}, \bibnamefont{and}
  \bibinfo{author}{\bibfnamefont{P.}~\bibnamefont{Lederer}},
  \bibinfo{journal}{Phys. Rev. B} \textbf{\bibinfo{volume}{68}},
  \bibinfo{pages}{205110} (\bibinfo{year}{2003}).

\bibitem[{\citenamefont{Rao and Sen}(2004)}]{Rao04}
\bibinfo{author}{\bibfnamefont{S.}~\bibnamefont{Rao}} \bibnamefont{and}
  \bibinfo{author}{\bibfnamefont{D.}~\bibnamefont{Sen}},
  \bibinfo{journal}{Phys. Rev. B} \textbf{\bibinfo{volume}{70}},
  \bibinfo{pages}{195115} (\bibinfo{year}{2004}).

\bibitem[{\citenamefont{Kazymyrenko and Dou\ifmmode~\mbox{\c{c}}\else
  \c{c}\fi{}ot}(2005)}]{Kazymyrenko05}
\bibinfo{author}{\bibfnamefont{K.}~\bibnamefont{Kazymyrenko}} \bibnamefont{and}
  \bibinfo{author}{\bibfnamefont{B.}~\bibnamefont{Dou\ifmmode~\mbox{\c{c}}\else
  \c{c}\fi{}ot}}, \bibinfo{journal}{Phys. Rev. B}
  \textbf{\bibinfo{volume}{71}}, \bibinfo{pages}{075110}
  (\bibinfo{year}{2005}).

\bibitem[{\citenamefont{Bellazzini et~al.}(2007)\citenamefont{Bellazzini,
  Mintchev, and Sorba}}]{Bellazzini07}
\bibinfo{author}{\bibfnamefont{B.}~\bibnamefont{Bellazzini}},
  \bibinfo{author}{\bibfnamefont{M.}~\bibnamefont{Mintchev}}, \bibnamefont{and}
  \bibinfo{author}{\bibfnamefont{P.}~\bibnamefont{Sorba}},
  \bibinfo{journal}{Journal of Physics A: Mathematical and Theoretical}
  \textbf{\bibinfo{volume}{40}}, \bibinfo{pages}{2485} (\bibinfo{year}{2007}).

\bibitem[{\citenamefont{Hou and Chamon}(2008)}]{Hou08}
\bibinfo{author}{\bibfnamefont{C.-Y.} \bibnamefont{Hou}} \bibnamefont{and}
  \bibinfo{author}{\bibfnamefont{C.}~\bibnamefont{Chamon}},
  \bibinfo{journal}{Phys. Rev. B} \textbf{\bibinfo{volume}{77}},
  \bibinfo{pages}{155422} (\bibinfo{year}{2008}).

\bibitem[{\citenamefont{Das and Rao}(2008)}]{Das08}
\bibinfo{author}{\bibfnamefont{S.}~\bibnamefont{Das}} \bibnamefont{and}
  \bibinfo{author}{\bibfnamefont{S.}~\bibnamefont{Rao}},
  \bibinfo{journal}{Phys. Rev. B} \textbf{\bibinfo{volume}{78}},
  \bibinfo{pages}{205421} (\bibinfo{year}{2008}).

\bibitem[{\citenamefont{Agarwal et~al.}(2009)\citenamefont{Agarwal, Das, Rao,
  and Sen}}]{Agarwal_Das_Rao_Sen09}
\bibinfo{author}{\bibfnamefont{A.}~\bibnamefont{Agarwal}},
  \bibinfo{author}{\bibfnamefont{S.}~\bibnamefont{Das}},
  \bibinfo{author}{\bibfnamefont{S.}~\bibnamefont{Rao}}, \bibnamefont{and}
  \bibinfo{author}{\bibfnamefont{D.}~\bibnamefont{Sen}},
  \bibinfo{journal}{Phys. Rev. Lett.} \textbf{\bibinfo{volume}{103}},
  \bibinfo{pages}{026401} (\bibinfo{year}{2009}).

\bibitem[{\citenamefont{Bellazzini
  et~al.}(2009{\natexlab{a}})\citenamefont{Bellazzini, Calabrese, and
  Mintchev}}]{Bellazzini09a}
\bibinfo{author}{\bibfnamefont{B.}~\bibnamefont{Bellazzini}},
  \bibinfo{author}{\bibfnamefont{P.}~\bibnamefont{Calabrese}},
  \bibnamefont{and} \bibinfo{author}{\bibfnamefont{M.}~\bibnamefont{Mintchev}},
  \bibinfo{journal}{Phys. Rev. B} \textbf{\bibinfo{volume}{79}},
  \bibinfo{pages}{085122} (\bibinfo{year}{2009}{\natexlab{a}}).

\bibitem[{\citenamefont{Bellazzini
  et~al.}(2009{\natexlab{b}})\citenamefont{Bellazzini, Mintchev, and
  Sorba}}]{Bellazzini09b}
\bibinfo{author}{\bibfnamefont{B.}~\bibnamefont{Bellazzini}},
  \bibinfo{author}{\bibfnamefont{M.}~\bibnamefont{Mintchev}}, \bibnamefont{and}
  \bibinfo{author}{\bibfnamefont{P.}~\bibnamefont{Sorba}},
  \bibinfo{journal}{Phys. Rev. B} \textbf{\bibinfo{volume}{80}},
  \bibinfo{pages}{245441} (\bibinfo{year}{2009}{\natexlab{b}}).

\bibitem[{\citenamefont{Safi}(2009)}]{Safi09}
\bibinfo{author}{\bibfnamefont{I.}~\bibnamefont{Safi}},
  \bibinfo{journal}{arXiv:0906.2363}  (\bibinfo{year}{2009}).

\bibitem[{\citenamefont{Aristov et~al.}(2010)\citenamefont{Aristov, Dmitriev,
  Gornyi, Kachorovskii, Polyakov, and W\"olfle}}]{Aristov10}
\bibinfo{author}{\bibfnamefont{D.~N.} \bibnamefont{Aristov}},
  \bibinfo{author}{\bibfnamefont{A.~P.} \bibnamefont{Dmitriev}},
  \bibinfo{author}{\bibfnamefont{I.~V.} \bibnamefont{Gornyi}},
  \bibinfo{author}{\bibfnamefont{V.~Y.} \bibnamefont{Kachorovskii}},
  \bibinfo{author}{\bibfnamefont{D.~G.} \bibnamefont{Polyakov}},
  \bibnamefont{and} \bibinfo{author}{\bibfnamefont{P.}~\bibnamefont{W\"olfle}},
  \bibinfo{journal}{Phys. Rev. Lett.} \textbf{\bibinfo{volume}{105}},
  \bibinfo{pages}{266404} (\bibinfo{year}{2010}).

\bibitem[{\citenamefont{Mintchev}(2011)}]{Mintchev11}
\bibinfo{author}{\bibfnamefont{M.}~\bibnamefont{Mintchev}},
  \bibinfo{journal}{Journal of Physics A: Mathematical and Theoretical}
  \textbf{\bibinfo{volume}{44}}, \bibinfo{pages}{415201}
  (\bibinfo{year}{2011}).

\bibitem[{\citenamefont{Aristov}(2011)}]{Aristov11}
\bibinfo{author}{\bibfnamefont{D.~N.} \bibnamefont{Aristov}},
  \bibinfo{journal}{Phys. Rev. B} \textbf{\bibinfo{volume}{83}},
  \bibinfo{pages}{115446} (\bibinfo{year}{2011}).

\bibitem[{\citenamefont{Aristov and W\"olfle}(2011)}]{Aristov_Wolfe11}
\bibinfo{author}{\bibfnamefont{D.~N.} \bibnamefont{Aristov}} \bibnamefont{and}
  \bibinfo{author}{\bibfnamefont{P.}~\bibnamefont{W\"olfle}},
  \bibinfo{journal}{Phys. Rev. B} \textbf{\bibinfo{volume}{84}},
  \bibinfo{pages}{155426} (\bibinfo{year}{2011}).

\bibitem[{\citenamefont{Wang and Feldman}(2011)}]{Wang_Feldman11}
\bibinfo{author}{\bibfnamefont{C.}~\bibnamefont{Wang}} \bibnamefont{and}
  \bibinfo{author}{\bibfnamefont{D.~E.} \bibnamefont{Feldman}},
  \bibinfo{journal}{Phys. Rev. B} \textbf{\bibinfo{volume}{83}},
  \bibinfo{pages}{045302} (\bibinfo{year}{2011}).

\bibitem[{\citenamefont{Caudrelier et~al.}(2012)\citenamefont{Caudrelier,
  Mintchev, and Ragoucy}}]{Caudrelier12}
\bibinfo{author}{\bibfnamefont{V.}~\bibnamefont{Caudrelier}},
  \bibinfo{author}{\bibfnamefont{M.}~\bibnamefont{Mintchev}}, \bibnamefont{and}
  \bibinfo{author}{\bibfnamefont{E.}~\bibnamefont{Ragoucy}},
  \bibinfo{journal}{arXiv:1202.4270}  (\bibinfo{year}{2012}).

\bibitem[{\citenamefont{Tomonaga}(1950)}]{Tomonaga50}
\bibinfo{author}{\bibfnamefont{S.}~\bibnamefont{Tomonaga}},
  \bibinfo{journal}{Prog. Theor. Phys.} \textbf{\bibinfo{volume}{5}},
  \bibinfo{pages}{544} (\bibinfo{year}{1950}).

\bibitem[{\citenamefont{Luttinger}(1963)}]{Luttinger63}
\bibinfo{author}{\bibfnamefont{J.~M.} \bibnamefont{Luttinger}},
  \bibinfo{journal}{J. Math. Phys.} \textbf{\bibinfo{volume}{4}},
  \bibinfo{pages}{1154} (\bibinfo{year}{1963}).

\bibitem[{\citenamefont{Mattis and Lieb}(1965)}]{Mattis65}
\bibinfo{author}{\bibfnamefont{D.~C.} \bibnamefont{Mattis}} \bibnamefont{and}
  \bibinfo{author}{\bibfnamefont{E.~H.} \bibnamefont{Lieb}},
  \bibinfo{journal}{J. Math. Phys.} \textbf{\bibinfo{volume}{6}},
  \bibinfo{pages}{304} (\bibinfo{year}{1965}).

\bibitem[{\citenamefont{Haldane}(1981)}]{Haldane81}
\bibinfo{author}{\bibfnamefont{F.~D.~M.} \bibnamefont{Haldane}},
  \bibinfo{journal}{J. Phys. C} \textbf{\bibinfo{volume}{14}},
  \bibinfo{pages}{2585} (\bibinfo{year}{1981}).

\bibitem[{\citenamefont{Hou et~al.}(2012)\citenamefont{Hou, Rahmani, Feiguin,
  and Chamon}}]{Hou12}
\bibinfo{author}{\bibfnamefont{C.-Y.} \bibnamefont{Hou}},
  \bibinfo{author}{\bibfnamefont{A.}~\bibnamefont{Rahmani}},
  \bibinfo{author}{\bibfnamefont{A.~E.} \bibnamefont{Feiguin}},
  \bibnamefont{and} \bibinfo{author}{\bibfnamefont{C.}~\bibnamefont{Chamon}},
  \bibinfo{journal}{Phys. Rev. B} \textbf{\bibinfo{volume}{86}},
  \bibinfo{pages}{075451} (\bibinfo{year}{2012}).

\end{thebibliography}

\end{document}